\documentclass{iopconfser}

\usepackage[T1]{fontenc}
\usepackage{lmodern}

\usepackage{amsthm}
\usepackage{amssymb}
\usepackage{amsmath}
\usepackage{mathtools}

\usepackage{adjustbox}

\usepackage{sidecap}

\usepackage
  [
    style=numeric-comp,
    sorting=none,
    backend=biber
  ]
  {biblatex}
\bibliography{references}

\usepackage{tikz}
\usetikzlibrary{
  cd,
  calc,
  arrows.meta,
  backgrounds,
  snakes
}

\tikzcdset{
  arrow style=tikz, 
  diagrams={
    >={
      Computer Modern Rightarrow[
        length=4pt, width=4pt
      ]
    },
    shorten=0pt
  }
}

\tikzset{
  snake left/.style={
    rounded corners,
    to path={
      let \p1 = (\tikztostart.east),
          \p2 = (\tikztotarget.west),
          \p3 = ($(\p1)!0.5!(\p2)$),
          \n1 = {8pt} 
      in
      (\p1)
      -- (\x1 + \n1, \y1)
      -- (\x1 + \n1, \y3)
      -- (\x2 - \n1, \y3) \tikztonodes
      -- (\x2 - \n1, \y2)
      -- (\p2)
    }
  }
}

\tikzset{
  uphordown/.style={
    rounded corners,
    to path={
      let \p1 = (\tikztostart.north),
          \p2 = (\tikztotarget.north),
          \n1 = {max(\y1,\y2) + 8pt}
      in
      (\p1)
      -- (\x1, \n1)
      -- (\x2, \n1) \tikztonodes 
      -- (\p2)
    }
  }
}

\tikzset{
  downhorup/.style={
    rounded corners,
    to path={
      let \p1 = (\tikztostart.south),
          \p2 = (\tikztotarget.south),
          \n1 = {min(\y1,\y2) - 8pt}
      in
      (\p1)
      -- (\x1, \n1)
      -- (\x2, \n1) \tikztonodes 
      -- (\p2)
    }
  }
}

\tikzset{
  rightvertleft/.style={
    rounded corners,
    to path={
      let \p1 = (\tikztostart.east),
          \p2 = (\tikztotarget.east),
          \n1 = {max(\x1,\x2) + 8pt}
      in
      (\p1)
      -- (\n1, \y1)
      -- (\n1, \y2) \tikztonodes 
      -- (\p2)
    }
  }
}

\tikzset{
  leftvertright/.style={
    rounded corners,
    to path={
      let \p1 = (\tikztostart.west),
          \p2 = (\tikztotarget.west),
          \n1 = {min(\x1,\x2) - 8pt}
      in
      (\p1)
      -- (\n1, \y1)
      -- (\n1, \y2) \tikztonodes 
      -- (\p2)
    }
  }
}

\definecolor{darkblue}{rgb}{0.05,0.25,0.65}
\definecolor{darkgreen}{RGB}{20,140,10}
\definecolor{lightgray}{rgb}{0.9,0.9,0.9}
\definecolor{darkorange}{RGB}{200,100,5}
\definecolor{darkyellow}{rgb}{.91,.91,0}
\definecolor{lightolive}{RGB}{225, 220, 185}

\usepackage[
  colorlinks=true,
  linkcolor=darkgreen,
  citecolor=darkgreen,
  urlcolor=darkgreen
]{hyperref}

\newcommand{\grayunderbrace}[2]{\mathcolor{gray}{\underbrace{\mathcolor{black}{#1}}}_{\mathcolor{gray}{#2}}}

\newcommand{\defneq}{\equiv}

\newcommand{\weakHomotopyEquivalence}{\sim}

\newcommand{\sslash}{\mathord{/\mkern-6mu/}}

\newcommand{\lattice}{K}

\newcommand{\p}{p}

\newcommand{\cpt}{\hspace{.8pt}{\adjustbox{scale={.5}{.77}}{$\cup$} \{\infty\}}}

\newcommand{\ActedIndexedOperator}[4]{          
       {#1}
          _{\hspace{-1.5pt}
           \scalebox{.6}{${#4}\hspace{-2pt}
           \left[
           \def\arraystretch{.8}
           \def\arraycolsep{0pt}
           \begin{array}{c}
            {#2} \\ #3
            \\[-12pt]
            {}
           \end{array}
           \right]
           $}
        }
}

\newcommand{\ActedWOperator}[3]{
  \ActedIndexedOperator{\widehat{W}}{#1}{#2}{#3}
}
\newcommand{\WOperator}[2]{
  \ActedWOperator{#1}{#2}{}
}

\newcommand{\hotype}[1]{\mathcal{#1}}

\begin{document}

%%%%%%%%%%%%%%%%%%%%%%%%%%%%%%%%%%%%%%%%%%
%vertical spacing around displayed equations %
\setlength{\abovedisplayskip}{2.2pt}
\setlength{\belowdisplayskip}{2.2pt}
\setlength{\abovedisplayshortskip}{-5pt}
\setlength{\belowdisplayshortskip}{2pt}
%%%%%%%%%%%%%%%%%%%%%%%%%%%%%%%%%%%%%%%

\title{
Non-Lagrangian Construction of Anyons
\\
\hspace{7pt}via Flux Quantization in Cohomotopy
}

\author{%
H. Sati$^{\ast}$ 
\;and\; 
U. Schreiber$^{\ast}$
}

\affil{
$^\ast$
Center for Quantum and Topological Systems, 
New York University, 
Abu Dhabi, 
UAE
}

\email{hsati@nyu.edu, us13@nyu.edu}

\begin{abstract}
We provide a brief invitation to the novel understanding 
\cite{SS25-AbelianAnyons, SS25-FQH, SS25-FQAH, SS25-Srni}
of anyonic topological order in fractional quantum (anomalous) Hall systems, via ``extraordinary'' quantization of effective magnetic flux in Cohomotopy --- following our presentation at ISQS29 \cite{SS25-ISQS29}.
\end{abstract}

\tableofcontents

%%%%%%%%%%%%%%%%%%%%%%%%%%%%%%%%%%%%%%%%%%
\section{Motivation and Introduction}
%%%%%%%%%%%%%%%%%%%%%%%%%%%%%%%%%%%%%%%%%
While {\it anyons} have been considered as a theoretically possible curiosity since the 1970s (\cite{Leinaas1977}, cf. \cite{Greiter2024}), it is the ongoing ``second quantum revolution'' (\cite{Dowling2003}, cf. \cite{nLabQuantumTechnology}) which promotes them --- in the guise of anyonic \emph{solitons} and \emph{defects} 
\cite{Mermin1979, SS23-Ord}
in topological quantum materials \cite{nLabQuantumMaterial}
--- to a tangible reality, whence their experimental realization and technological potential
warrant a more proper theoretical understanding by mathematical physicists, which, despite the decades of discussion, has arguably remained sketchy.

\paragraph{Topological quantum and Anyons.}
It is hard to overstate the hopes \cite{Gill2025} associated with the idea of digital quantum technology/computing \cite{Nielsen2012,nLabQuantumTechnology}. And yet the core problem remains essentially unsolved \cite{Waintal2023}: The stabilization of quantum registers against decohering noise that jeopardizes the creation of large-scale robust quantum computers of practical value.  

Apart from the popular approach of {\it quantum error correction} \cite{Lidar2013} by means of heavy software-level redundancy, this stabilization plausibly necessitates (cf. \cite{Freedman2002, DasSarma2022}) {\it error protection}, already at the hardware level, by physically suppressing decoherence in the first place. The ambitious idea of {\it topological quantum computing} (\cite{Kitaev2003}, cf. \cite{Freedman2002, SV25-TQC}) is to employ {\it topological quantum states} as quantum registers, whose coherence is largely protected by fundamental physical principles. 

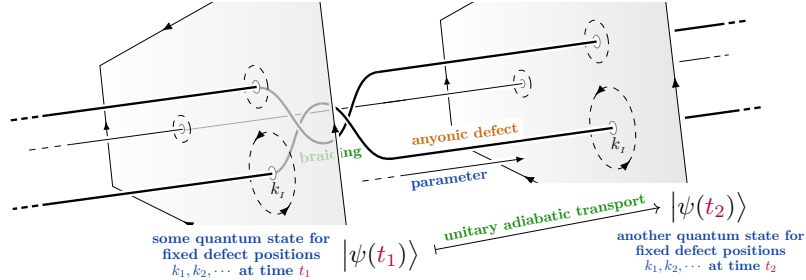
\begin{figure}
\centering
\caption{
  \label{TopologicalQuantumGate}
  {Topological quantum gates} by adiabatic braiding of anyon worldlines. 
}
\vspace{2mm} 
\adjustbox{
  scale=.8,
  raise=0mm,
}{
\begin{tikzpicture}[
  baseline=(current bounding box.center)
]

\clip
  (-8.4, 1.3) rectangle
  (4.9,-4.7);

  \shade[right color=lightgray, left color=white]
    (3,-3)
      --
    (-1,-1)
      --
    (-1.21,1)
      --
    (2.3,3);

  \draw[]
    (3,-3)
      --
    (-1,-1)
      --
    (-1.21,1)
      --
    (2.3,3)
      --
    (3,-3);

\draw[-Latex]
  ({-1 + (3+1)*.3},{-1+(-3+1)*.3})
    to
  ({-1 + (3+1)*.29},{-1+(-3+1)*.29});

\draw[-Latex]
    ({-1.21 + (2.3+1.21)*.3},{1+(3-1)*.3})
      --
    ({-1.21 + (2.3+1.21)*.29},{1+(3-1)*.29});

\draw[-Latex]
    ({2.3 + (3-2.3)*.5},{3+(-3-3)*.5})
      --
    ({2.3 + (3-2.3)*.49},{3+(-3-3)*.49});

\draw[-latex]
    ({-1 + (-1.21+1)*.53},{-1 + (1+1)*.53})
      --
    ({-1 + (-1.21+1)*.54},{-1 + (1+1)*.54});

  \begin{scope}[rotate=(+8)]
   \draw[dashed]
     (1.5,-1)
     ellipse
     ({.2*1.85} and {.37*1.85});
   \begin{scope}[
     shift={(1.5-.2,{-1+.37*1.85-.1})}
   ]
     \draw[->, -Latex]
       (0,0)
       to
       (180+37:0.01);
   \end{scope}
   \begin{scope}[
     shift={(1.5+.2,{-1-.37*1.85+.1})}
   ]
     \draw[->, -Latex]
       (0,0)
       to
       (+37:0.01);
   \end{scope}
   \begin{scope}[shift={(1.5,-1)}]
     \draw (.43,.65) node
     { \scalebox{.8}{$
     $} };
  \end{scope}
  \draw[fill=white, draw=gray]
    (1.5,-1)
    ellipse
    ({.2*.3} and {.37*.3});
  \draw[line width=3.5, white]
   (1.5,-1)
   to
   (-2.2,-1);
  \draw[line width=1.1]
   (1.5,-1)
   to node[
     above, 
     yshift=-4pt, 
     pos=.85]{
     \;\;\;\;\;\;\;\;\;\;\;\;\;
     \rotatebox[origin=c]{7}
     {
     \scalebox{.7}{
     \color{darkorange}
     \bf
     \colorbox{white}{anyonic defect}
     }
     }
   }
   (-2.2,-1);
  \draw[
    line width=1.1
  ]
   (1.5+1.2,-1)
   to
   (3.5,-1);
  \draw[
    line width=1.1,
    densely dashed
  ]
   (3.5,-1)
   to
   (4,-1);

  \draw[line width=3, white]
   (-2,-1.3)
   to
   (0,-1.3);
  \draw[-latex]
   (-2,-1.3)
   to
   node[
     below, 
     yshift=+3pt,
     xshift=-7pt
    ]{
     \scalebox{.7}{
       \rotatebox{+7}{
       \color{darkblue}
       \bf
       parameter
       }
     }
   }
   (0,-1.3);
  \draw[dashed]
   (-2.7,-1.3)
   to
   (-2,-1.3);

 \draw
   (-3.15,-.8)
   node{
     \scalebox{.7}{
       \rotatebox{+7}{
       \color{darkgreen}
       \bf
       braiding
       }
     }
   };

  \end{scope}

  \begin{scope}[shift={(-.2,1.4)}, scale=(.96)]
  \begin{scope}[rotate=(+8)]
  \draw[dashed]
    (1.5,-1)
    ellipse
    (.2 and .37);
  \draw[fill=white, draw=gray]
    (1.5,-1)
    ellipse
    ({.2*.3} and {.37*.3});
  \draw[line width=3.1, white]
   (1.5,-1)
   to
   (-2.3,-1);
  \draw[line width=1.1]
   (1.5,-1)
   to
   (-2.3,-1);
  \draw[line width=1.1]
   (1.5+1.35,-1)
   to
   (3.6,-1);
  \draw[
    line width=1.1,
    densely dashed
  ]
   (3.6,-1)
   to
   (4.1,-1);
  \end{scope}
  \end{scope}

  \begin{scope}[shift={(-1,.5)}, scale=(.7)]
  \begin{scope}[rotate=(+8)]
  \draw[dashed]
    (1.5,-1)
    ellipse
    (.2 and .32);
  \draw[fill=white, draw=gray]
    (1.5,-1)
    ellipse
    ({.2*.3} and {.32*.3});
  \draw[line width=3.1, white]
   (1.5,-1)
   to
   (-1.8,-1);
\draw
   (1.5,-1)
   to
   (-1.8,-1);
  \draw
    (5.23,-1)
    to
    (6.4-.6,-1);
  \draw[densely dashed]
    (6.4-.6,-1)
    to
    (6.4,-1);
  \end{scope}
  \end{scope}

\draw (1.73,-1.06) node
 {
  \scalebox{.8}{
    $k_{{}_{I}}$
  }
 };

\begin{scope}
[ shift={(-2,-.55)}, rotate=-82.2  ]

 \begin{scope}[shift={(0,-.15)}]

  \draw[]
    (-.2,.4)
    to
    (-.2,-2);

  \draw[
    white,
    line width=1.1+1.9
  ]
    (-.73,0)
    .. controls (-.73,-.5) and (+.73-.4,-.5) ..
    (+.73-.4,-1);
  \draw[
    line width=1.1
  ]
    (-.73+.01,0)
    .. controls (-.73+.01,-.5) and (+.73-.4,-.5) ..
    (+.73-.4,-1);

  \draw[
    white,
    line width=1.1+1.9
  ]
    (+.73-.1,0)
    .. controls (+.73,-.5) and (-.73+.4,-.5) ..
    (-.73+.4,-1);
  \draw[
    line width=1.1
  ]
    (+.73,0+.03)
    .. controls (+.73,-.5) and (-.73+.4,-.5) ..
    (-.73+.4,-1);

  \draw[
    line width=1.1+1.9,
    white
  ]
    (-.73+.4,-1)
    .. controls (-.73+.4,-1.5) and (+.73,-1.5) ..
    (+.73,-2);
  \draw[
    line width=1.1
  ]
    (-.73+.4,-1)
    .. controls (-.73+.4,-1.5) and (+.73,-1.5) ..
    (+.73,-2);

  \draw[
    white,
    line width=1.1+1.9
  ]
    (+.73-.4,-1)
    .. controls (+.73-.4,-1.5) and (-.73,-1.5) ..
    (-.73,-2);
  \draw[
    line width=1.1
  ]
    (+.73-.4,-1)
    .. controls (+.73-.4,-1.5) and (-.73,-1.5) ..
    (-.73,-2);

 \draw
   (-.2,-3.3)
   to
   (-.2,-2);
 \draw[
   line width=1.1,
   densely dashed
 ]
   (-.73,-2)
   to
   (-.73,-2.5);
 \draw[
   line width=1.1,
   densely dashed
 ]
   (+.73,-2)
   to
   (+.73,-2.5);

  \end{scope}
\end{scope}

\begin{scope}[shift={(-5.6,-.75)}]

  \draw[line width=3pt, white]
    (3,-3)
      --
    (-1,-1)
      --
    (-1.21,1)
      --
    (2.3,3)
      --
    (3, -3);

  \shade[right color=lightgray, left color=white, fill opacity=.7]
    (3,-3)
      --
    (-1,-1)
      --
    (-1.21,1)
      --
    (2.3,3);

  \draw[]
    (3,-3)
      --
    (-1,-1)
      --
    (-1.21,1)
      --
    (2.3,3)
      --
    (3, -3);

\draw (1.73,-1.06) node
 {
  \scalebox{.8}{
    $k_{{}_{I}}$
  }
 };

\draw[-Latex]
  ({-1 + (3+1)*.3},{-1+(-3+1)*.3})
    to
  ({-1 + (3+1)*.29},{-1+(-3+1)*.29});

\draw[-Latex]
    ({-1.21 + (2.3+1.21)*.3},{1+(3-1)*.3})
      --
    ({-1.21 + (2.3+1.21)*.29},{1+(3-1)*.29});

\draw[-Latex]
    ({2.3 + (3-2.3)*.5},{3+(-3-3)*.5})
      --
    ({2.3 + (3-2.3)*.49},{3+(-3-3)*.49});

\draw[-latex]
    ({-1 + (-1.21+1)*.53},{-1 + (1+1)*.53})
      --
    ({-1 + (-1.21+1)*.54},{-1 + (1+1)*.54});

  \begin{scope}[rotate=(+8)]
   \draw[dashed]
     (1.5,-1)
     ellipse
     ({.2*1.85} and {.37*1.85});
   \begin{scope}[
     shift={(1.5-.2,{-1+.37*1.85-.1})}
   ]
     \draw[->, -Latex]
       (0,0)
       to
       (180+37:0.01);
   \end{scope}
   \begin{scope}[
     shift={(1.5+.2,{-1-.37*1.85+.1})}
   ]
     \draw[->, -Latex]
       (0,0)
       to
       (+37:0.01);
   \end{scope}
  \draw[fill=white, draw=gray]
    (1.5,-1)
    ellipse
    ({.2*.3} and {.37*.3});
 \end{scope}

   \begin{scope}[shift={(-.2,1.4)}, scale=(.96)]
  \begin{scope}[rotate=(+8)]
  \draw[dashed]
    (1.5,-1)
    ellipse
    (.2 and .37);
  \draw[fill=white, draw=gray]
    (1.5,-1)
    ellipse
    ({.2*.3} and {.37*.3});
\end{scope}
\end{scope}

  \begin{scope}[shift={(-1,.5)}, scale=(.7)]
  \begin{scope}[rotate=(+8)]
  \draw[dashed]
    (1.5,-1)
    ellipse
    (.2 and .32);
  \draw[fill=white, draw=gray]
    (1.5,-1)
    ellipse
    ({.2*.3} and {.37*.3});
\end{scope}
\end{scope}

\begin{scope}
[ shift={(-2,-.55)}, rotate=-82.2  ]

 \begin{scope}[shift={(0,-.15)}]

 \draw[line width=3, white]
   (-.2,-.2)
   to
   (-.2,2.35);
 \draw
   (-.2,.5)
   to
   (-.2,2.35);
 \draw[dashed]
   (-.2,-.2)
   to
   (-.2,.5);

\end{scope}
\end{scope}

\begin{scope}
[ shift={(-2,-.55)}, rotate=-82.2  ]

 \begin{scope}[shift={(0,-.15)}]

 \draw[
   line width=3, white
 ]
   (-.73,-.5)
   to
   (-.73,3.65);
 \draw[
   line width=1.1
 ]
   (-.73,.2)
   to
   (-.73,3.65);
 \draw[
   line width=1.1,
   densely dashed
 ]
   (-.73,.2)
   to
   (-.73,-.5);
 \end{scope}
 \end{scope}

\begin{scope}
[ shift={(-2,-.55)}, rotate=-82.2  ]

 \begin{scope}[shift={(0,-.15)}]

 \draw[
   line width=3.2,
   white]
   (+.73,-.6)
   to
   (+.73,+3.7);
 \draw[
   line width=1.1,
   densely dashed]
   (+.73,-0)
   to
   (+.73,+-.6);
 \draw[
   line width=1.1 ]
   (+.73,-0)
   to
   (+.73,+3.71);
\end{scope}
\end{scope}

\end{scope}

\draw[
  draw=white,
  fill=white
]
  (-8,-2.4) rectangle
  (3,-3.8);

\draw[
  draw=white,
  fill=white
]
  (-1,-1.7) rectangle
  (3.4,-2.5);

\begin{scope}[
  shift={(0,1.3)}
]
\draw
  (-2.2,-4.2) node
  {
    \adjustbox{
      bgcolor=white,
      scale=1.2
    }{
      $
       \mathllap{
          \raisebox{1pt}{
            \scalebox{.58}{
              \color{darkblue}
              \bf
              \def\arraystretch{.9}
              \begin{tabular}{c}
                some quantum state for
                \\
                fixed defect positions
                \\
                $k_1, k_2, \cdots$
                at time
                {\color{purple}$t_1$}
              \end{tabular}
            }
          }
          \hspace{-5pt}
       }
        \big\vert
          \psi({\color{purple}t_1})
        \big\rangle
      $
    }
  };

\draw
  (+3.2,-3.85) node
  {
    \adjustbox{
      bgcolor=white,
      scale=1.2
    }{
      $
        \underset{
          \raisebox{-7pt}{
            \scalebox{.55}{
              \color{darkblue}
              \bf
              \def\arraystretch{.9}
               \begin{tabular}{c}
              another quantum state for
                \\
                fixed defect positions
                \\
                $k_1, k_2, \cdots$
                at time
                {\color{purple}$t_2$}
              \end{tabular}
            }
          }
        }{
        \big\vert
          \psi({\color{purple}t_2})
        \big\rangle
        }
      $
    }
  };

\draw[|->]
  (-1.3,-4.1)
  to
  node[
    sloped,
    yshift=5pt
  ]{
    \scalebox{.7}{
      \color{darkgreen}
      \bf
      unitary adiabatic transport
    }
  }
  node[
    sloped,
    yshift=-5pt,
    pos=.4
  ]{
    \scalebox{.7}{
      }
  }
  (+2.4,-3.4);
\end{scope}

\end{tikzpicture}
}
\vspace{-1.2cm}

\end{figure}

The prominent example, in theory, are effectively 2-dimensional quantum materials exibiting {\it topological order} \cite{nLabTopologicalOrder}, where the (adiabatic) movement of anyonic soliton/defect positions in the material effects unitary transformations of its ground states which only depend on the isotopy class of the deformation path. In this way 
the {\it braiding} of worldlines of such {\it anyonic} solitons/defects 
may serve as {\it topological quantum gates} insensitive to local noise (cf. \cite[\S 3]{MySS2024} and see  Fig. \ref{TopologicalQuantumGate}).

%%%%%%%%%%%%%%%%%%%%%%%%%%%%%%%%%%
\section{Solitonic Anyons in FQH Systems}
\label{SolitonicAnyons}
%%%%%%%%%%%%%%%%%%%%%%%%%%%%%%%%%%

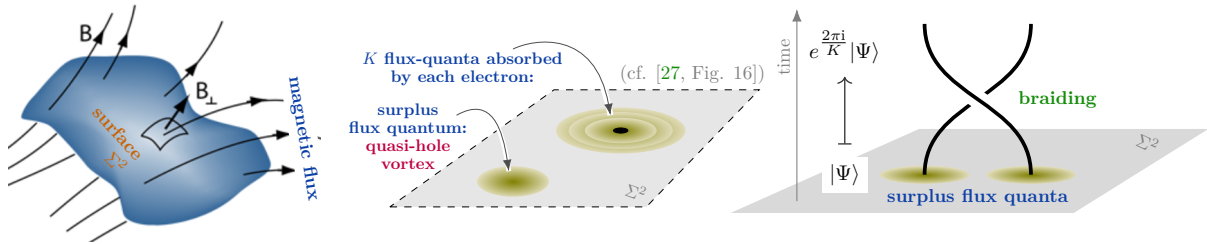
\begin{figure}[htb]
\caption{
\label{Flux}
{Anyons in FQH systems} are (quasi-hole vortices associated with) surplus flux magnetic flux quanta 
(relative to a given rational \emph{filling fraction} of $\lattice$ flux quanta per electron)
through an electron gas occupying an effectively 2-dimensional semiconducting surface $\Sigma^2$. This suggests \cite{SS25-FQH} that FQH anyons are to be understood in terms of an exotic effective {\it flux quantization law} \cite{SS25-Flux}.}

\adjustbox{
  raise=-1.9cm,
  scale=1
}{
\begin{tikzpicture}
\node at (0,0)
{
\includegraphics[width=4.1cm]{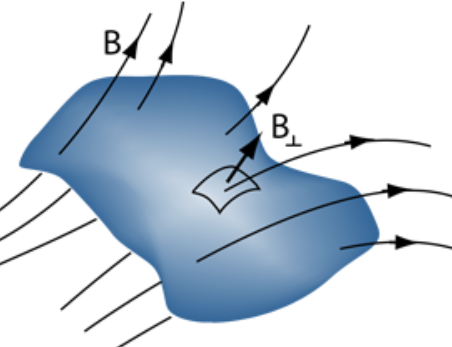}
};

\node[
  scale=.7,
  rotate=-81
] at (1.83,-.2) {
  \colorbox{white}{%
  \color{darkblue}%
  \bf%
  magnetic flux%
  }
};

\node[
  scale=.7,
  rotate=-48
] at (-.74,-.24) {
  \color{darkorange}
  \bf
  \begin{tabular}{c}
    surface 
    \\
    \;\;\;\;\;\;\;$\Sigma^2$
  \end{tabular}
};
\end{tikzpicture}
}
\hspace{-.6cm}
\adjustbox{
  raise=-1.5cm,
  scale=.9
}{
\begin{tikzpicture}[
  scale=.75
]

\node
  at (.3,.55+.8)
  {
    \adjustbox{
      bgcolor=white,
      scale=.7
    }{
      \color{darkblue}
      \bf
      \def\arraystretch{.9}
      \begin{tabular}{c}
        surplus
        \\
        flux quantum:
        \\
        \color{purple}
        quasi-hole
        \\
        \color{purple}
        vortex
      \end{tabular}
    }
  };

\draw[
  dashed,
  fill=lightgray
]
  (0,0)
  -- (5,0)
  -- (7+.3-.1,2+.3)
  -- (2.8+.3+.1,2+.3)
  -- cycle;

\begin{scope}[
  shift={(2.4,.5)}
]
\shadedraw[
  draw opacity=0,
  inner color=olive,
  outer color=lightolive
]
  (0,0) ellipse (.7 and .3);
\end{scope}

\begin{scope}[
  shift={(4.5,1.5)}
]

\begin{scope}[
 scale=1.8
]
\shadedraw[
  draw opacity=0,
  inner color=olive,
  outer color=lightolive
]
  (0,0) ellipse (.7 and .25);
\end{scope}

\begin{scope}[
 scale=1.45
]
\shadedraw[
  draw opacity=0,
  inner color=olive,
  outer color=lightolive
]
  (0,0) ellipse (.7 and .25);
\end{scope}

\shadedraw[
  draw opacity=0,
  inner color=olive,
  outer color=lightolive
]
  (0,0) ellipse (.7 and .25);

\begin{scope}[
  scale=.2
]
\draw[
  fill=black
]
  (0,0) ellipse (.7 and .25);
\end{scope}

\end{scope}

\draw[
  white,
  line width=2
]
  (1.3, 1.8) .. controls 
  (2,2.2) and 
  (2.2,1.5) ..
  (2.32,.7);
\draw[
  -Latex,
  black!70
]
  (1.3, 1.8) .. controls 
  (2,2.2) and 
  (2.2,1.5) ..
  (2.32,.7);

\node
  at (1.3,2.7)
  {
    \adjustbox{
      scale=.7
    }{
      \color{darkblue}
      \bf
      \def\arraystretch{.9}
      \def\tabcolsep{-5pt}
      \begin{tabular}{c}
        $\lattice$ flux-quanta
        absorbed
        \\
        by each electron:
      \end{tabular}
    }
  };

\draw[
 line width=2.5pt,
  white
]
  (2.4, 3.1) .. controls 
  (2.8,3.3) and 
  (4,3.5) ..
  (4.3,1.8);

\draw[
  -Latex,
  black!70
]
  (2.4, 3.1) .. controls 
  (2.8,3.3) and 
  (4,3.5) ..
  (4.3,1.8);

\node at 
  (5.9,2.6)
  {
   \scalebox{.8}{
     \color{gray}
     (cf. \cite[Fig. 16]{Stormer99})  
   }
  };

\node[
  gray,
  rotate=-20,
  scale=.73
] 
  at (4.8,+.3) {$\Sigma^2$};

\end{tikzpicture}
}
\hspace{-1cm}
  \adjustbox{raise=-1.4cm}{
  \begin{tikzpicture}[
    xscale=.7
  ]
    \draw[
      gray!30,
      fill=gray!30
    ]
      (-4.6,-1.5) --
      (+1.8,-1.5) --
      (+1.8+3-.5,-.4) --
      (-4.6+3+.5,-.4) -- cycle;

    \begin{scope}[
      shift={(-1,-1)},
      scale=1.2
    ]
    \shadedraw[
      draw opacity=0,
      inner color=olive,
      outer color=lightolive
    ]
      (0,0) ellipse (.7 and .1);
    \end{scope}

    \draw[
     line width=1.4
    ]
      (-1,-1) .. controls
      (-1,0) and
      (+1,0) ..
      (+1,+1);

  \begin{scope}
    \clip 
      (-1.5,-.2) rectangle (+1.5,1);
    \draw[
     line width=7,
     white
    ]
      (+1,-1) .. controls
      (+1,0) and
      (-1,0) ..
      (-1,+1);
  \end{scope}
  
    \begin{scope}[
      shift={(+1,-1)},
      scale=1.2
    ]
    \shadedraw[
      draw opacity=0,
      inner color=olive,
      outer color=lightolive
    ]
      (0,0) ellipse (.7 and .1);
    \end{scope}
    \draw[
     line width=1.4
    ]
      (+1,-1) .. controls
      (+1,0) and
      (-1,0) ..
      (-1,+1);

  \node[
    rotate=-25,
    scale=.7,
    gray
  ]
    at (3.2,-.58) {
      $\Sigma^2$
    };

  \draw[
    -Latex,
    gray
  ]
    (-3.4,-1.35) -- 
    node[
      near end, 
      sloped,
      scale=.7,
      yshift=7pt
      ] {time}
    (-3.4, 1.2);

  \node[
    scale=.7
  ] at 
    (0,-1.3)
   {\bf \color{darkblue} 
   surplus flux quanta};

  \node[
    scale=.7
  ] at 
    (1.5,0)
   {\bf \color{darkgreen} braiding};

  \node[
    fill=white,
    scale=.8
  ] at (-2.5,-1) {$
    \vert \Psi \rangle
  $};

  \node[
    fill=white,
    scale=.8
  ] at (-2.5,.7) {$
    e^{\tfrac{2 \pi \mathrm{i}}{K}}
    \vert \Psi \rangle
  $};

  \draw[
    |->,
    black!80,
    line width=.5
  ]
    (-2.5,-.6) --
    (-2.5, .3);

  \end{tikzpicture}
  }
\vspace{-.6cm}

\end{figure}

Despite considerable interest in this basic idea since its proposal over 25 years ago, the theoretical understanding of real anyonic quantum materials had remained sketchy (cf. \cite[\S 5.1]{Jain2007}\cite[\S 1]{Halperin2020}) and its experimental realization elusive (cf. \cite{DasSarma2023, nLab:MajoranaZeroMode}). 
That is, until recently: 

\paragraph{Fractional Quantum Hall Anyons.}
On the experimental side, in the last years the observation of braiding phases of (abelian) anyons in {\it fractional quantum Hall} systems (FQH, cf. \cite{Stormer99, Jain2007, Tong2016}) have come to be consistently reported by several groups and in various materials \cite{nLab:AnyonsInFQH}. On the theoretical side, we survey here a new rigorous understanding 
\cite{SS25-AbelianAnyons, SS25-FQH, SS25-FQAH, SS25-Srni}
of such FQH anyons by means of the previously overlooked issue of their proper {\it flux quantization} 
\footnote{
  Here \emph{flux quantization}  (also: ``charge quantization'') refers, a priori, to ``quantization'' in the sense of \emph{discretization}: The traditional \emph{Dirac charge quantization} of ordinary electromagnetism \cite{Alvarez1985}\cite[\S 16.4e]{Frankel2011} famously constrains electric charges to be integer multiples of a unit quantum of charge, and also constrains, in particular, solitonic magnetic flux through the plane  to come in integer units of the magnetic \emph{flux quantum} \cite[\S 2.1]{SS25-Flux}, as indicated in Fig. \ref{Flux}. In more detail, Dirac charge quantization constrains the magnetic field to a class in ordinary integral 2-cohomology of spacetime, cf. \eqref{CohomologyAndCohomotopy} below.
  However, flux/charge quantization is closely related to actual \emph{quantization} in the sense of quantum mechanics/field theory: Historically, Dirac charge quantization is necessary for the quantum mechanics of electrically charged particles to be globally consistent. More importantly for our purpose, the choice of flux quantization of a(n effective, higher, ...) gauge field immediately determines the \emph{topological} flux quantum observables of the corresponding quantum fields, hence the quantum theory of the topological sector of the field -- this is the main result of \cite{SS24-Obs}, used in \eqref{OrdinaryFluxOnTorus} and \eqref{AnyonAlgebraOverTorus} below, following \cite[\S 2.1]{SS25-FQH}
} (cf. \cite{SS24-Phase, SS25-Flux}).

For this we highlight (cf. \cite[p. 882]{Stormer99} and see Figs. \ref{Flux} and \ref{SuperconductingIslands}) that the characteristic aspect of the anyons (observed) specifically in FQH systems is that they {\it are flux quanta} of the magnetic field, or rather are their imprint, in the guise of {\it vortices}, onto the strongly correlated electron gas that constitutes the 2D FQH system.

\paragraph{Flux quantization -- Ordinary and exotic.}
In the absence of strongly-correlated electrons, away from their exact fractional Landau level filling, standard electromagnetic theory predicts (\cite{SS24-Obs}, via ordinary ``Dirac flux quantization'' \cite{Alvarez1985}\cite[\S 2.1]{SS25-Flux}) that their topological quantum observables are entirely controlled by the algebraic topology (the homotopy theory) of the ordinary {\it classifying space} $B \mathrm{U}(1) \,\weakHomotopyEquivalence\, \mathbb{C}P^\infty$ of $\mathrm{U}(1)$ gauge field, cf. \cite{SS25-Flux}, in that the topological flux quantum observables form this commutative algebra \cite[(15)]{SS25-FQH}:
\begin{equation}
  \label{OrdinaryFluxOnTorus}
  \begin{aligned}
  \mathbb{C}
  \Big[
  \pi_1
  \,
  \mathrm{Map}^\ast\big(
    \Sigma^2_{\cpt}
    ,\,
    \mathcolor{purple}{\mathbb{C}P^\infty}
  \big)
  \Big]
  &
  \simeq
  \mathbb{C}
  \Big[
  \pi_0
  \,
  \mathrm{Map}^\ast\big(
    \Sigma^2_{\cpt}
    ,\,
    S^1
  \big)
  \Big]
  \,\simeq\,
  \mathbb{C}\Big[
    \widetilde{H}^1\big(
      \Sigma^2_{\cpt}
      ;\,
      \mathbb{Z}
    \big)
  \Big]
  \\
  &
  \underset{\mathclap{%
    \Sigma^2 = T^2
  }}{
    \simeq
  }
  \mathbb{C}\Big[
    \WOperator{1}{0}
    ,
    \WOperator{0}{1}
  \Big]
  \Big/
  \Big(
    \WOperator{1}{0}
    \WOperator{0}{1}
    =
    \WOperator{0}{1}
    \WOperator{1}{0}
  \Big)
  \,.
  \end{aligned}
\end{equation}

\vspace{1mm} 
However, since this prediction is crucially violated for FQH flux quanta imprinted on vortices in the 2D electron gas (Fig. \ref{Flux}), while flux quantization as such must clearly remain in effect in some form, we conclude with \cite{SS25-FQH} that the situation is to be described by a variant {\it effective} flux quantization law for exotic FQH flux. Since flux quantization laws are determined by their classifying spaces \cite[\S 3.2]{SS25-Flux}\cite{FSS23-Char}, we are to look for a suitable variant $\hotype{A}$ of the classifying space $\mathbb{C}P^\infty$. 

This $\hotype{A}$ will be a choice determining our effective model of the physics, just like the familiar specification of any effective Lagrangian density is a choice of a physics model. To the extent that such a choice implies known characteristic properties of the physical systems we tend to trust it and regard its further implications as predictions about previously unknown properties of the physical system.
We consider the simplest choice, in a sense: The first \emph{skeleton} of $\mathbb{C}P^\infty$, which the (2-)sphere:
\begin{equation}
  S^2
  \simeq
  \mathbb{C}P^1 
  \xhookrightarrow{\;\;}
  \textstyle{\bigcup_{n \in \mathbb{N}}}
  \,
  \mathbb{C}P^n
  \simeq
  \mathbb{C}P^\infty
  \weakHomotopyEquivalence
  B \mathrm{U}(1)
  \,.
\end{equation}
Where $\mathbb{C}P^\infty$ is the classifying space for \emph{ordinary} integral 2-cohomology $\widetilde{H}(-;\mathbb{Z})$ (cf. \cite[Ex. 2.1]{FSS23-Char}), so $S^2 \simeq \mathbb{C}P^1$ is the classifying space for the unstable/nonabelian \emph{extraordinary} cohomology theory known as \emph{2-Cohomotopy} $\pi^2(-)$ (cf. \cite[\S 7]{STHu1959}\cite[Ex. 2.7]{FSS23-Char}):
\begin{equation}
  \label{CohomologyAndCohomotopy}
  \begin{aligned}
  \widetilde{H}^2\big(
    -;
    \mathbb{Z}
  \big)
  &
  \simeq
  \pi_0
  \big(
  \mathrm{Map}^\ast(
   -
   ,
   \mathbb{C}P^\infty
  )
  \big)
  \;
  \adjustbox{
    scale=.7
  }{
    \color{gray}
    \bf
    ordinary cohomology
  }
  \\
  \widetilde{\pi}^2(-)
  &
  \defneq
  \pi_0
  \big(
  \mathrm{Map}^\ast(
   -
   ,
   \mathbb{C}P^{1\;}
  )
  \big)
  \;
  \adjustbox{
    scale=.7
  }{
    \color{gray}
    \bf
    extraordinary cohomotopy
  }
  \end{aligned}
\end{equation}
Therefore we shall refer to the hypothesis that $\mathbb{C}P^1$ is the correct classifying space for surplus FQH flux as \emph{Hypothesis h} (a small cousin \cite{SS25-Seifert, SS25-Srni}
of the capital \emph{Hypothesis H} in high energy physics \cite{FSS20-H}\cite{FSS21-Hopf}).

\smallskip 
The first phenomenological justification for \emph{Hypothesis h} is the following most remarkable fact: Over the torus,
it yields the following
noncommutative modification of \eqref{OrdinaryFluxOnTorus} --- by \cite[Prop. 3.19]{SS25-FQH}, using \cite[Thm. 1]{LarmoreThomas1980}, cf. \cite[Prop. 1.5]{Kallel2001}, going back to \cite{Hansen1974}:
\begin{equation}
  \label{AnyonAlgebraOverTorus}
  \begin{aligned}
  \mathbb{C}
  \Big[
  \pi_1
  \,
  \mathrm{Map}^\ast\big(
    \Sigma^2_{\cpt}
    ,\,
    \mathcolor{purple}{\mathbb{C}P^1}
  \big)
  \Big]
  &
  \simeq
  \mathbb{C}
  \Big[
  \pi_0
  \,
  \mathrm{Map}^\ast\big(
    \Sigma^2_{\cpt}
    ,\,
    S^1
  \big)
  \Big]
  \\
  &
  \underset{\mathclap{%
    \Sigma^2 = T^2
  }}{
    \simeq
  }
  \mathbb{C}\Big[
    \WOperator{1}{0}
    ,
    \WOperator{0}{1}
    ,
    \underset{
     \mathrm{central}
    }{
      \mathcolor{purple}{\zeta}
    }
  \Big]
  \Big/
  \Big(
    \WOperator{1}{0}
    \WOperator{0}{1}
    =
    \mathcolor{purple}{\zeta^2}
    \,
    \WOperator{0}{1}
    \WOperator{1}{0}
  \Big)
  \,.
  \end{aligned}
\end{equation}
But this is just algebra of abelian Chern-Simons \emph{Wilson loop observables} on the torus, thought to characterize FQH anyons there (cf. \cite[(5.28)]{Tong2016})!

\paragraph{Relation to Chern-Simons theory.}
Indeed, traditionally the end result of \eqref{AnyonAlgebraOverTorus} is argued 
instead 
(as recalled in \cite[\S A.1]{SS25-FQH})
by first arguing that a form of the abelian Chern-Simons Lagrangian density \cite{nLabAbelianCS} provides the \emph{effective field theory} for FQH anyons. But as highlighted in \cite[Rem. A.1]{SS25-FQH}, this traditional assumption appears to be at odds with ordinary flux quantization (as noted in \cite[p. 35]{Witten2016}\cite[p. 159]{Tong2016}), which is worrying in view of the very flux quantum nature of FQH anyons (Fig. \ref{Flux}). Indeed, at generic filling fraction the Chern-Simons model of FQH anyons is an elaboration of the \emph{Haldane-Halperin hierarchy model}, which is known to be unphysical for most filling fractions (see \cite[Rem. 2.1]{nLabAbelianCS}).

\vspace{1mm} 
In our novel description of FQH anyons this situation is turned right-side-up: Proper flux quantization is made the very starting point of the description of anyonic FQH flux quanta, and Lagrangian densities (which generally do not reflect global topological properties of fields, cf. \cite[Fig. G]{SS25-FQH}) are never used or needed.
Remarkably, the resulting non-Lagrangian theory of FQH anyons broadly agrees with abelian Chern-Simons theory, but makes distinct experimentally discernible predictions for certain filling fractions (see \cite[p. 5]{SS25-FQH}).

\paragraph{Anyon worldlines and their braiding.}
More tangibly, flux quantization in 2-cohomotopy manifests the worldlines of anyonic solitons and their braiding phases, as follows.

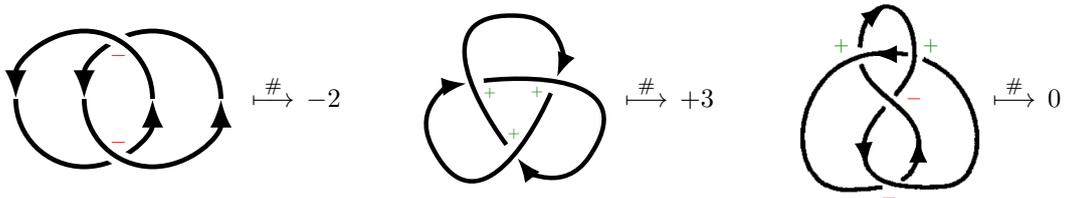
\begin{figure}[htb]
\centering
\caption{
  \label{FramedLinks}
  Some (blackboard-)framed Wilson loop/links and their total crossing number/writhe. 
}
$
\adjustbox{
  raise=-1cm,
  scale=.9
}{
\begin{tikzpicture}[
  scale=1
]

\begin{scope}[
  shift={(1,0)}
]
\draw[line width=2, -Latex]
  (0:1) arc (0:180:1);
\end{scope}

\draw[line width=7,white]
  (0:1) arc (0:180:1);
\draw[line width=2, -Latex]
  (0:1) arc (0:180:1);

\draw[line width=2, -Latex]
  (180:1) arc (180:360:1);

\begin{scope}[shift={(1,0)}]
\draw[line width=7, white]
  (180:1) arc (180:360:1);
\draw[line width=2, -Latex]
  (180:1) arc (180:360:1);
\end{scope}

\node[gray]
  at (.5,.64) {\color{red} 
    \scalebox{.9}{$-$}
  };
\node[gray]
  at (.5,-.64) {\color{red}
    \scalebox{.9}{$-$}
  };
  
\end{tikzpicture}
}
\overset{\#}{\longmapsto}
-2
$
\hspace{.6cm}
$
\adjustbox{
  raise=-2.7cm,
  scale=.6
}{
\begin{tikzpicture}
\foreach \n in {0,1,2} {
\begin{scope}[
  rotate=\n*120-4
]
\draw[
  line width=2.8,
  -Latex
]
 (0-.1,-1+.14)
   .. controls
   (-1,.2) and (-2,2) ..
 (0,2)
   .. controls
   (1,2) and (1,1) ..
  (.9,.7);
\end{scope}

\node[darkgreen]
  at (\n*120+31:.6) {
    \scalebox{1}{$+$}
  };

};
\end{tikzpicture}
\hspace{-25pt}
}
\overset{\#}{\longmapsto}
+3
$
\hspace{.6cm}
$
\adjustbox{
  raise=-1.3cm,
  scale=1.1
}{
\begin{tikzpicture}

\node at (0,0) {
  \includegraphics[width=2.2cm]{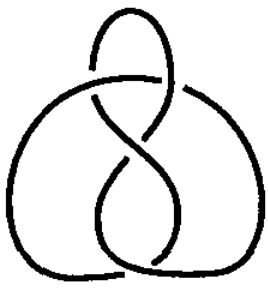}
};

\draw[line width=1.2pt,-Latex]
  (-.276,-.8) -- 
  (-.274,-.8-.02);
\draw[line width=1.2pt,-Latex]
  (.35, -.5) --
  (.35, -.47);
\draw[line width=1.2pt,-Latex]
  (-.09,1.09) --
  (-.09+.05, 1.13);
\draw[line width=1.2pt,-Latex]
  (-.18,.54) --
  (-.19-.01, .54);

\node[red]
  at (0,-1.2) {
    \scalebox{.7}{$-$}
  };
\node[red]
  at (.3,0) {
    \scalebox{.7}{$-$}
  };
\node[darkgreen]
  at (.5,.64) {
    \scalebox{.7}{$+$}
  };
\node[darkgreen]
  at (-.57,.64) {
    \scalebox{.7}{$+$}
  };

\end{tikzpicture}
\hspace{-10pt}
}
\overset{\#}{\longmapsto} 0
$

\vspace{-.8cm}

\end{figure}

Looking at the case of flux through the plane, $\Sigma^2 \defneq \mathbb{R}^2$ --- which is the situation naturally realized in FQH experiments ---the topological quantum observables according to \emph{Hypothesis h} are
\begin{equation}
  \label{QuantumObservablesOnThePlane}
  \mathbb{C}\Big[
  \pi_1 \,
  \mathrm{Map}^\ast\big(
    \mathbb{R}^2_{\cpt}
    ,\,
    \mathcolor{purple}{\mathbb{C}P^1}
  \big)
  \Big]
  \simeq
  \mathbb{C}\big[
  \pi_0 \,
  \mathrm{Map}^\ast\big(
    S^3
    ,\,
    S^2
  \big)
  \big]
  \defneq
  \mathbb{C}\big[
    \widetilde{\pi}^2(S^3)
  \big]
  \simeq
  \mathbb{C}\big[
    \mathbb{Z}
  \big]
  \,,
\end{equation}
where the last identification is via the \emph{Hopf fibration} \cite{Lyons2003}
\begin{equation}
  \big[S^3 \xrightarrow{h} S^2\big]
  =
  1
  \in
  \mathbb{Z}
  \simeq
  \pi^2(S^3)
  =
  \pi_3(S^2)
  \,.
\end{equation}
Now,  \emph{Pontrjagin's theorem} 
\cite{nLabPontrjagin}
identifies $\widetilde{\pi}^2(S^3)$
with the cobordism classes 
$\mathrm{Cob}^2_{\mathrm{fr}}(S^3)$
of 
\emph{framed links} (cf. \cite[\S2]{SS25-AbelianAnyons} and see Fig. \ref{FramedLinks}), hence with regularized \emph{Wilson loops} $L$, and careful analysis shows
\cite[Thm. 2.19]{SS25-AbelianAnyons}
that their integer characteristic on the right of \eqref{QuantumObservablesOnThePlane} is their \emph{total crossing number} (or \emph{writhe}) $\# L$:
\begin{equation}
  \begin{tikzcd}[sep=0pt]
  \mathrm{Cob}^2_{\mathrm{fr}}(S^3)
  \ar[
    rr, 
    "{ \sim }"
  ]
  &&
  \widetilde \pi^2(S^3)
  \simeq
  \mathbb{Z}
  \\
  {[L]} &\longmapsto& \# L
  \end{tikzcd}
\end{equation}

It follows \cite[Cor. 3.3]{SS25-AbelianAnyons} that the topological pure quantum states (of 2-cohomotopical flux through the plane) are labeled by a number
\begin{equation}
  \label{UnquantizedLevel}
  K \in \mathbb{R}\setminus \{0\}
\end{equation}
and that the expectation values of framed Wilson loops/link observables in these states are the exponentiated sum of framing and linking numbers:
\begin{equation}
  \begin{tikzcd}[
    column sep=0pt,
    row sep=-3pt
  ]
    \mathbb{C}\big[
      \mathrm{Cob}^2_{\mathrm{fr}}(S^3)
    \big]
    \ar[
      rr,
      "{
        \langle K \vert - \vert K \rangle
      }"
    ]
    &&
    \mathbb{C}
    &
    \\
    {[L]} &\longmapsto&
    \exp\big(
      \tfrac{2 \pi \mathrm{i}}
      {K}
      \# L
    \big)
    &=&
    \exp\Big(
      \tfrac{2 \pi \mathrm{i}}
      {K}
      \big(
        \sum_i
        \mathrm{frm}(L_i)
        +
        \sum_{i,j} 
        \mathrm{lnk}(L_i, L_j)
      \big)
    \Big)
    \mathrlap{\,.}
  \end{tikzcd}
\end{equation}
On the right we recognize the traditional Wilson loop observable of abelian Chern-Simons theory (\cite[Rem. 3.4]{SS25-AbelianAnyons}, whose regularization by framing is traditionally an \emph{ad hoc} fix of an ill-defined path integral quantization procedure \cite[Rem. 3.5]{SS25-AbelianAnyons}, while here it is a rigorous consequence of \emph{Hypothesis h}, cf. \cite[Rem. 3.6]{SS25-AbelianAnyons}), whereas in the middle this is recognized as assigning a fixed \emph{braiding phase} 
$e^{\tfrac{2\pi \mathrm{i}}{K}}$ to each crossing/braiding, just as it should be for FQH anyons (by Fig. \ref{Flux}).

\paragraph{General covariance and Topological order}
More precisely, a \emph{topological field theory}, like that of FQH anyons, is to be ``generally covariant'' --- in that field/flux configurations which differ only by a diffeomorphism of the domain are to be regarded as gauge-equivalent. In terms of classifying spaces $\hotype{A}$ this means that the moduli space of topological flux is not exactly the plain mapping space from $\Sigma^2$ to $\hotype{A}$, but is its \emph{homotopy quotient} $(-)\sslash \mathrm{Diff}(\Sigma^2)$ (``Borel construction'') by the diffeomorphism group of $\Sigma^2$ \cite[(21)]{SS25-FQH}. Therefore the \emph{covariantized topological quantum observables} are, more precisely, the group algebra of the semidirect product of the flux monodromy group with the \emph{mapping class group} $\mathrm{MCG}$ \cite[Prop. 2.24]{SS25-FQH}:
\begin{equation}
 \label{GenerallyCovariantTopologicalQuantumObservables}
 \grayunderbrace{
   \mathrm{TQObs}_{\Sigma^2}^{\hotype{A}}
 }{
   \mathclap{
     \scalebox{.7}{
       \def\arraystretch{.9}
       \begin{tabular}{c}
         topological flux
         \\
         quantum observables
       \end{tabular}
     }
   }
 }
 \coloneqq
 \mathbb{C}
 \bigg[
 \grayunderbrace{
 \pi_1
   \Big(
     \mathrm{Map}^\ast\big(
       \Sigma^2_{\cpt}
       ,\,
       \hotype{A}
     \big)
     \mathcolor{purple}{
       \sslash
       \mathrm{Diff}(\Sigma^2)
     }
   \Big)
   }{%
    \scalebox{.7}{%
      generally covariant flux monodromy%
    }
   }
 \bigg]
 \simeq
 \mathbb{C}
 \bigg[
   \Big(
    \grayunderbrace{
    \pi_1
     \mathrm{Map}^\ast\big(
       \Sigma^2_{\cpt}
       ,\,
       \hotype{A}
     \big)
     }{
       \scalebox{.7}{
         flux monodromy
       }
     }
    \Big)
    \mathcolor{purple}{
      \rtimes
      \grayunderbrace{
      \mathcolor{purple}{%
        \pi_0 \mathrm{Diff}(\Sigma^2)%
      }
      }{
        \mathrm{MCG}(\Sigma^2)
      }
    }
 \bigg]
 \,,
\end{equation}

Over the torus, $\Sigma^2 \defneq T^2$, the mapping class group is the \emph{modular group}, and we find 
\cite[\S 3.4]{SS25-FQH} that for flux quantization in 2-cohomotopy $\hotype{A} \defneq S^2$ \eqref{CohomologyAndCohomotopy}, the general covariance of \eqref{GenerallyCovariantTopologicalQuantumObservables} enforces the \emph{modular data} of Chern-Simons/Wess-Zumino-Witten theory \cite[Rem. 3.39]{SS25-FQH}, in particular it implies --- on top of the anyonic observables from \eqref{AnyonAlgebraOverTorus} --- that on superselection sectors (on quantum state spaces which are irreps of the observable algebra) the level \eqref{UnquantizedLevel} is quantized as appropriate for spin Chern-Simons theory, and that the braiding phases (Fig. \ref{Flux}) are roots of unity, as seen in FQH systems. The resulting irreducible modules of $\mathrm{TQObs}_{T^2}^{S^2}$ reflect the detailed anyonic \emph{topological order} predicted by Hypothesis h, which turns out to broadly agree with traditional arguments form abelian Chern-Simons theory, but adds some fine print and seems to differ in some aspects for some filling fractions \cite[p. 5 \& Rem. 3.45]{SS25-FQH}.

%%%%%%%%%%%%%%%%%%%%%%%%%%%%%%%%%%%%%%%%
\section{Defect Anyons in FQH Systems}
%%%%%%%%%%%%%%%%%%%%%%%%%%%%%%%%%%%%%%%%

\begin{figure}[htb]
\centering
\caption{
  \label{SuperconductingIslands}
  Where FQH anyons are solitonic quanta of \emph{concentrations} of magnetic flux (cf. Fig. \ref{Flux}), super-conducting island in the semi-conductor substrate will tend to \emph{expel} magnetic flux. \emph{Hypothesis h} implies/predicts that, if adiabatically movable, such islands behave like \emph{defect anyons}.
}

\begin{tikzpicture}[scale=0.8]

\draw[
  gray!30,
  fill=gray!15
]
  (0-.3,0) --
  (-.3+3,1) --
  (9.7+3, 1) --
  (9.7, 0) -- cycle;

\foreach \n in {1,...,13} {
  \draw[
    line width=2pt
  ]
    (\n*.7, -2) --
    (\n*.7, -1);
  \draw[
    line width=2pt
  ]
    (\n*.7, +1) --
    (\n*.7, +2);
}

\draw[
  line width=2pt
]
  (4*.7, -1) .. controls
  (4*.7, -.3) and
  (4*.7+.3,-.6) ..
  (4*.7+.3, 0);
\draw[
  line width=2pt
]
  (4*.7, +1) .. controls
  (4*.7, +.3) and
  (4*.7+.3,+.6) ..
  (4*.7+.3, 0);

\draw[
  line width=2pt
]
  (3*.7, -1) .. controls
  (3*.7, -.3) and
  (3*.7-.3,-.6) ..
  (3*.7-.3, 0);
\draw[
  line width=2pt
]
  (3*.7, +1) .. controls
  (3*.7, +.3) and
  (3*.7-.3,+.6) ..
  (3*.7-.3, 0);

\draw[
  line width=2pt
]
  (11*.7, -1) .. controls
  (11*.7, -.3) and
  (11*.7+.2,-.6) ..
  (11*.7+.2, 0);
\draw[
  line width=2pt
]
  (11*.7, +1) .. controls
  (11*.7, +.3) and
  (11*.7+.2,+.6) ..
  (11*.7+.2, 0);

\draw[
  line width=2pt
]
  (10*.7, -1) .. controls
  (10*.7, -.3) and
  (10*.7-.2,-.6) ..
  (10*.7-.2, 0);
\draw[
  line width=2pt
]
  (10*.7, +1) .. controls
  (10*.7, +.3) and
  (10*.7-.2,+.6) ..
  (10*.7-.2, 0);

\draw[
  line width=2pt
]
  (6*.7, -1) .. controls
  (6*.7, -.3) and
  (6*.7+.75,-.6) ..
  (6*.7+.75, 0);

\draw[
  line width=2pt
]
  (9*.7, -1) .. controls
  (9*.7, -.3) and
  (9*.7-.75,-.6) ..
  (9*.7-.75, 0);

\draw[
  line width=2pt
]
  (7*.7, -1) .. controls
  (7*.7, -.3) and
  (7*.7+.25,-.6) ..
  (7*.7+.25, 0);
\draw[
  line width=2pt
]
  (8*.7, -1) .. controls
  (8*.7, -.3) and
  (8*.7-.25,-.6) ..
  (8*.7-.25, 0);

\draw[
  gray,
  line width=2
]
  (-.3,0) --
  (2.05,0);
\draw[
  gray,
  line width=2
]
  (2.85,0) --
  (6.95,0);
\draw[
  gray,
  line width=2
]
  (7.75,0) --
  (9.7,0);
  
\shadedraw[
  draw opacity=0,
  inner color=olive,
  outer color=lightolive
]
  (7.5*.7,0) ellipse 
  (.58 and .12);

\draw[
  line width=2pt
]
  (7*.7, +1) .. controls
  (7*.7, +.3) and
  (7*.7+.25,+.6) ..
  (7*.7+.25, 0);

\draw[
  line width=2pt
]
  (8*.7, +1) .. controls
  (8*.7, +.3) and
  (8*.7-.25,+.6) ..
  (8*.7-.25, 0);

\draw[
  line width=2pt
]
  (6*.7, +1) .. controls
  (6*.7, +.3) and
  (6*.7+.75,+.6) ..
  (6*.7+.75, 0);

\draw[
  line width=2pt
]
  (9*.7, +1) .. controls
  (9*.7, +.3) and
  (9*.7-.75,+.6) ..
  (9*.7-.75, 0);

\draw[
  line width=2pt
]
  (1*.7, -1) --
  (1*.7, +1); 

\draw[
  line width=2pt
]
  (2*.7, -1) .. controls
  (2*.7, -.3) and
  (2*.7-.1,-.6) ..
  (2*.7-.1, 0);
\draw[
  line width=2pt
]
  (2*.7, +1) .. controls
  (2*.7, +.3) and
  (2*.7-.1,+.6) ..
  (2*.7-.1, 0);

\draw[
  line width=2pt
]
  (5*.7, -1) .. controls
  (5*.7, -.3) and
  (5*.7+.1,-.6) ..
  (5*.7+.1, 0);
\draw[
  line width=2pt
]
  (5*.7, +1) .. controls
  (5*.7, +.3) and
  (5*.7+.1,+.6) ..
  (5*.7+.1, 0);

\draw[
  line width=2pt
]
  (12*.7, -1) --
  (12*.7, +1); 

\draw[
  line width=2pt
]
  (13*.7, -1) --
  (13*.7, +1); 

\draw[
  gray!30,
  draw opacity=.5,
  fill=gray!25,
  fill opacity=.5
]
  (0-.3,0) --
  (-.3-3,-1) --
  (9.7-3, -1) --
  (9.7,0) -- cycle;

\draw[fill=white] 
  (3.5*.7,0) ellipse 
  (.57 and .07);
\draw[fill=white] 
  (10.5*.7,0) ellipse 
  (.4 and .07);

\begin{scope}

\clip
  (4,0) rectangle
  (+6,-1);

\shadedraw[
  draw opacity=0,
  inner color=olive,
  outer color=lightolive
]
  (7.5*.7,0) ellipse 
  (.58 and .12);

\end{scope}

\node at (2.4,-2.5)
 {
   \scalebox{.65}{
     \color{darkblue}
     \bf
     \def\arraystretch{.9}
     \begin{tabular}{c}
       flux-expelling
       \\
       defect (puncture):
       \\
       {\bf non-abelian} anyon
     \end{tabular}
   }
 };

\node at (5.3,-2.5)
 {
   \scalebox{.65}{
     \color{darkblue}
     \bf
     \def\arraystretch{.9}
     \begin{tabular}{c}
       flux quantum
       \\
       soliton (vortex):
       \\
     abelian anyon
     \end{tabular}
     }
 };

\node[
  scale=.8,
  rotate=-25,
  gray
] at (-1.3,-.7) {$\Sigma^2$};

\end{tikzpicture}
\vspace{-.5cm}

\end{figure}
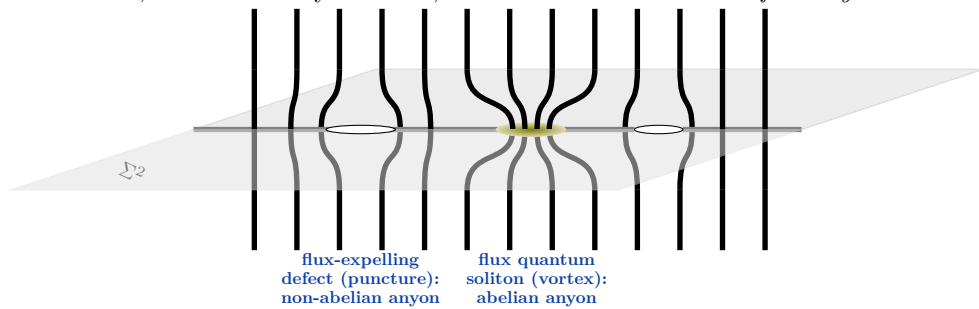

Results \cite{SS25-FQH} as indicated above in \S\ref{SolitonicAnyons} show the \emph{Hypothesis h} --- that surplus FQH flux is quantized in 2-Cohomotopy --- reproduces key properties observed or expected for FQH anyons. 
It is then interesting to note that the hypothesis makes further predictions of phenomena that have not (or cannot) be discussed by traditional means, but that may be visible experimentally and be relevant technologically. We highlight one of these:

\paragraph{Superconducting Islands in FQH Systems.}
The idea that topological quantum states might be realized in super-/semi-conducting heterostructures has received a tremendous amount of attention for 1D materials (``Majorana zero modes'') but persistently so with at best inconclusive experimental verification \cite{DasSarma2023, nLab:MajoranaZeroMode}. Instead, we may consider the doping of 2D semiconducting FQH systems by super-conducting islands, which, due to the Meissner effect, will tend to expel the magnetic flux (see Fig. \ref{SuperconductingIslands}).

Now, the evident mathematical model for the presence of a finite set $D \subset \Sigma^2$ of such flux expelling defect points is 
\cite[Fig. D \& \S3.5]{SS25-FQH}
to replace the surface $\Sigma^2$ with the \emph{punctured surface} $\Sigma^2 \setminus D$. Indeed, as the topological phases of flux are labeled by pointed maps
$
  \mathrm{Map}^\ast\big(
    (\Sigma^2 \setminus D)_{\cpt}
    ,
    \hotype{A}
  \big)
$
out of the one-point compactification of the domain, this encodes the constraint of vanishing flux not only at literal infinity --- as was the case in \eqref{QuantumObservablesOnThePlane} --- but also at each of the defect points in $D$, since in $(\Sigma^2 \setminus D)_{\cpt}$ all of these are regarded as being ``at infinity'' where the pointedness of the classifying maps forces flux to vanish.

But this has the remarkable consequence that with such flux-expelling defects included, the covariantized topological quantum observables \eqref{GenerallyCovariantTopologicalQuantumObservables} reflect operations of the surface \emph{braid group}
(cf. \cite[\S2.2]{SS25-FQH})
\begin{equation}
  \begin{tikzcd}
  \mathrm{Br}_{\vert D \vert}(\Sigma^2)
  \ar[r]
  &
  \mathrm{MCG}\big(
    \Sigma^2 \setminus D
  \big)  
  \mathrlap{\,,}
  \end{tikzcd}
\end{equation}
whose $\vert D \vert \in \mathbb{N}$ strands are the wordlines of these very flux-expelling defects. This implies that the corresponding space of topological quantum states of the FQH system are predicted to constitute a \emph{braid representation} \cite[\S3.5,8]{SS25-FQH}, where each braiding of worldlines of defects acts as a unitary transformation as in Fig. \ref{TopologicalQuantumGate}, witnessing the flux-expelling defects as \emph{defect anyons}.

Moreover --- in contrast to the \emph{solitonic anyons} that arose from the flux monodromy in \S\ref{SolitonicAnyons} and that have been experimentally observed in FQH experiments --- the theory does not constrain these defect anyons to be abelian.

%(...)

%%%%%%%%%%%%%%%%%%%%%%%%%%%%%%%%%%%%%%%
\section{Solitonic Anyons in FQAH Systems}
\label{SolitonicAnyonsInFQAH}
%%%%%%%%%%%%%%%%%%%%%%%%%%%%%%%%%%%%%%%

Very recently an ``anomalous'' version of the fractional quantum Hall effect (FQAH, cf. \cite{nLab:QAHE}) has been observed in \emph{fractional Chern insulators} (FCI): Crystalline quantum materials in which the role of the magnetic flux in physical space is instead played by \emph{Berry curvature} over the \emph{Brillouin torus} $\widehat{\mathbb{T}^2}$ of crystal quasi-momenta \cite{nLab:BrillouinTorus}. This suggests the tantalizing possibility that FAQH systems may serve as topological quantum hardware under much more practical laboratory conditions than FQH systems --- but the nature and appearance of their anyon braiding operation, if any, has not been explained by traditional theory.

\begin{SCfigure}[][htb]
\caption{
  \label{FQAHSystem}
  Fractional Chern insulators are $\sim2D$ crystalline materials exhibiting a fractional quantum \emph{anomalous} Hall effect (FQAH), where the role of physical space is played by the ``reciprocal'' space of crystal momenta, the \emph{Brillouin torus} $\widehat{T}^2$, and the role of magnetic flux is played by the \emph{Berry curvature} of the valence bundle of electron Bloch quantum states at given momentum. With Berry curvature concentrations around band maxima, these again correspond to electron holes, but now localized in momentum space. 
}

\adjustbox{
  scale=0.81
}{
\begin{tikzpicture}[
    >=Stealth,
]

\draw[
  line width=2,
  darkblue
]
  (-.5,-2) .. controls
  (3,-2) and
  (2,-.5) ..
  (3, -.5) .. controls
  (3.5,-.5) and
  (3.5,-2) ..
  (6.5,-2);

\begin{scope}[
  shift={(0,-.7)},
  yscale=-1
]
\draw[
  line width=2,
  darkblue
]
  (-.5,-2) .. controls
  (3,-2) and
  (2,-.5) ..
  (3, -.5) .. controls
  (3.5,-.5) and
  (3.5,-2) ..
  (6.5,-2);
\end{scope}

\draw[
  draw opacity=0,
  fill=white,
  fill opacity=.7
]
  (-.5,-1) rectangle
  (6.5,1.5);
  
\draw[
  ->,
  gray,
  line width=1.5
]
  (0,-3.5) --
    node[
      pos=.8,
      xshift=-3pt
    ] {
      \llap{
        Energy 
        $\epsilon$
      }
    }
  (0,2);

\draw[
  ->,
  gray,
  line width=1.5
]
  (-1.2,-3) --
    node[
      pos=1.1,
      yshift=-14pt
    ] {
      \llap{
        \def\arraystretch{.7}
        \begin{tabular}{c}
          $k_x$
          \\
          Momentum 
        \end{tabular}
      }
    }
  (6.8,-3);

\draw[
  ->,
  gray,
  line width=1.5,
  shift={(0,-3)}
]
  (16:-1.4cm) --
    node[
      pos=.85,
      shift={(-2pt,+3pt)}
    ]{
      \colorbox{white}{$k_y$}
    }
  (16:+3.4cm);

\draw[
  line width=1,
  color=purple,
  dashed
]
  (-.1,-1) --
  (6.5,-1);
\node[
  gray
] at  (-0.05,-1) {
  \llap{
  Fermi $\epsilon_{{}_{F}}$
  }
};

\node[scale=.7] at 
  (6,-2.4) {
    \def\arraystretch{1}
    \begin{tabular}{c}
      Valence band
      \\
      (Chern number$=+C$)
    \end{tabular}
  };

\node[scale=.7] at 
  (5.7,+.8) {
    \def\arraystretch{1}
    \begin{tabular}{c}
      Conduction band
      \\
      (Chern number $= -C$)
    \end{tabular}
  };

\node[
  scale=.8
] at (3.8,-.3) {
  \def\arraystretch{1}
  \begin{tabular}{c}
    gapped 
    \\
    Dirac cone
  \end{tabular}
};

\node[
  scale=.9
] at 
  (2.95,-1.25) {hole};

\shadedraw[
  draw opacity=0,
  inner color=olive,
  outer color=lightolive!50,
  shift={(2.93,-3)},
  scale=.65
]
  (0,0) ellipse (1.8 and .3);

\node[
  scale=.9
] at (3,-3.7) {
  \def\arraystretch{.9}
  \begin{tabular}{c}
    Berry curvature 
    \\
    concentration
  \end{tabular}
};

\node[
 scale=.9,
 rotate=-40,
 gray
] 
  at (1.54,-2.8) 
  {$\widehat{T}^2$};

\end{tikzpicture}
}

\end{SCfigure}
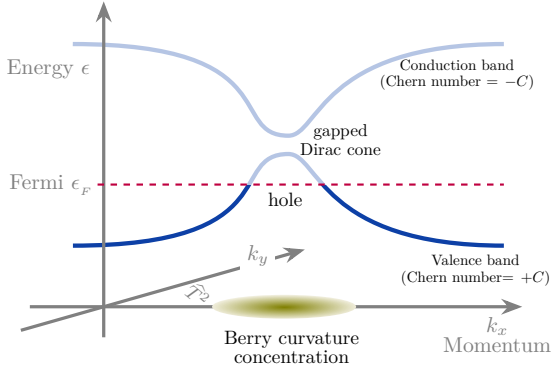

On the other hand, it is well-known that the topological phases of typical FCI are classified by homotopy classes of maps from the Brillouin torus to a 2-sphere of relative normalized 2-band Bloch Hamiltonians (cf. \cite[(8.3-8.4)]{Sergeev2023}):
\begin{equation}
  \begin{tikzcd}[column sep=huge]
    \widehat{T}^2
    \ar[
      rrrr,
      uphordown,
      "{
        H / \vert H \vert 
      }",
      "{
        \scalebox{.7}{
          \bf
          \color{darkgreen}
          normalized Bloch Hamiltonians
        }
      }"{swap}
    ]
    \ar[
      rr,
      dashed,
      "{
        \scalebox{.7}{
          \color{darkgreen}
          \bf
          \def\arraystretch{.9}
          \begin{tabular}{c}
            given by map
            \\
            to 2-sphere
          \end{tabular}
        }
      }"{swap}
    ]
    &&
    S^2
    \ar[
      rr,
      "{
        \vec x 
        \,\mapsto\,
        x^i \sigma_i
      }",
      "{
        \scalebox{.7}{%
          \color{gray}%
          Pauli matrices%
        }
      }"{swap}
    ]
    &&
    \mathrm{Mat}_{2 \times 2}(\mathbb{C})
    \mathrlap{\,.}
  \end{tikzcd}
\end{equation}
This means, we may observe \cite{SS25-FQAH}, that the \emph{fragile band topology} (\cite{Bouhon2020}, meaning: not coarsened by deformations through further electron conduction bands) is quantized in 2-Cohomotopy \eqref{CohomologyAndCohomotopy}! While this observation makes no difference for the topological (Chern) class, due to the \emph{Hopf degree theorem}:
\begin{equation}
  \pi_0
  \,\mathrm{Map}^\ast\big(
    \widehat{\mathbb{T}}^2_{\cpt}
    ,\,
    \mathbb{C}P^1
  \big)
  \defneq
  \widetilde{\pi}^2\big(
    \widehat{\mathbb{T}}^2
  \big)
  \;\;
  \underset{
    \mathclap{%
    \adjustbox{
      scale=.7,
      raise=-3pt
    }{%
      \color{gray}%
      Hopf degree
    }
    }
  }{
    \simeq
  }
  \;\;
  \widetilde{H}^2\big(
    \widehat{\mathbb{T}}^2
    ;\,
    \mathbb{Z}
  \big)
  \simeq
  \mathbb{Z}
  \mathrlap{\,,}
\end{equation}
and hence may and has been ignored for that purpose,
we observe that it does make a key difference for the potentially anyonic \emph{monodromy} in the space of topological phases \cite{SS25-FQAH}: As in \eqref{AnyonAlgebraOverTorus}, this is anyonic:
\begin{equation}
  \mathbb{C}
  \Big[
  \pi_1\, 
  \mathrm{Map}\big(
    \widehat{\mathbb{T}}^2
    ,\,
    \mathbb{C}P^1
  \big)
  \Big]
  \simeq
  \mathbb{C}\Big[
    \WOperator{1}{0}
    ,
    \WOperator{0}{1}
    ,
    \underset{
     \mathrm{central}
    }{
      \mathcolor{purple}{\zeta}
    }
  \Big]
  \Big/
  \Big(
    \WOperator{1}{0}
    \WOperator{0}{1}
    =
    \mathcolor{purple}{\zeta^2}
    \,
    \WOperator{0}{1}
    \WOperator{1}{0}
  \Big)  
  \,.
\end{equation}
This analysis predicts anyonic topological order in 2-band FCI which break all crystalline symmetries, with anyons localized not in ordinary physical space but in the ``reciprocal'' space $\widehat{\mathbb{T}}^2$ of crystal momenta (previously considered also in \cite{SS23-Ord}). Remarkably, the toroidal topology which makes this work --- which for FQH systems in ordinary space is experimentally at best hard to approximate --- is the default topology of the crystal's momentum space: the Brillouin torus \cite{nLab:BrillouinTorus}!  In this way, anyonic topological order
over the momentum space of FCI may be the most natural experimental realization of this phenomenon.

%%%%%%%%%%%%%%%%%%%%%%%%%%%%%%%%%%%%%%%
\section{Conclusion and Outlook}
%%%%%%%%%%%%%%%%%%%%%%%%%%%%%%%%%%%%%%

Despite their long history of discussion and their more recent practical relevance as potential topological quantum hardware, the actual theoretical understanding of anyonic solitons/defects in quantum materials has remained sketchy. In view of the fact that precisely one kind of such anyons has been unambiguously experimentally observed in recent years --- FQH anyons --- and highlighting the nature of these as (vortex quasi-holes associated with) \emph{flux quanta}, we reviewed an argument \cite[\S2]{SS25-FQH} that FQH anyons must be effectively modeled via a choice of ``extraordinary''  \emph{flux quantization law}, that, unlike traditional Lagrangian effective (Chern-Simons) field theory, properly captures their global topological properties. 

\smallskip 
The \emph{Hypothesis h} that the appropriate choice of effective FQH flux quantization is in the extraordinary nonabelian/unstable cohomology theory known as (2-)\emph{Cohomotopy} turns out to imply at once \cite{SS25-AbelianAnyons}\cite[\S3]{SS25-FQH} the hallmark properties observed of solitonic FQH anyons or expected via Chern-Simons arguments: fractional statistics, topological order and edge modes. Furthermore, \emph{Hypothesis h} provably predicts that (superconducting) flux-expelling islands in FQH semiconductor materials constitute possibly non-abelian \emph{defect anyons} if adiabatically movable --- which would be dramatic if experimentally verified. 

\smallskip 
Finally, we highlighted with \cite{SS25-FQAH} that \emph{Hypothesis h} is actually \emph{known} to apply to the \emph{fragile band topology} of generic crystalline FQAH systems --- which have very recently been experimentally observed, but whose anyonic nature had remained elusive --- predicting that FQAH anons exist as fragile band monodromy effects localized in reciprocal momentum space.

%%%%%%%%%%%%%%%%%%%%%%%%
  \printbibliography
  % \bibliographystyle{iopart-num}
  % \bibliography{references}
%%%%%%%%%%%%%%%%%%%%%%%

\end{document}